\documentclass[journal=jacsat,manuscript=article]{achemso}

\usepackage{chemformula} % Formula subscripts using \ch{}
\usepackage[T1]{fontenc} % Use modern font encodings
\usepackage[table]{xcolor} % Enable \rowcolor for tables
\usepackage{amsmath}
\usepackage{breqn}

\usepackage{microtype} 
\author{Li Liang}
\affiliation[Unknown University]
{Key Laboratory of Quantum Materials and Devices of Ministry of Education, School of Physics, Southeast University,
Nanjing 211189, China}
\author{Ding Ning}
\affiliation[Unknown University]
{Key Laboratory of Quantum Materials and Devices of Ministry of Education, School of Physics, Southeast University,
Nanjing 211189, China}
\author{Mingqiang Gu}
\affiliation[Unknown University]
{School of Flexible Electronics, Sun Yat-sen University, Shenzhen 518107,  China}
\author{Shanshan Wang}
\affiliation[Unknown University]
{Key Laboratory of Quantum Materials and Devices of Ministry of Education, School of Physics, Southeast University,
Nanjing 211189, China}
\email{wangss@seu.edu.cn}
\author{Alessandro Stroppa}
\affiliation[Unknown University]
{CNR-SPIN (L’Aquila), Italy}
\email{alessandro.stroppa@spin.cnr.it}
\title[An \textsf{achemso} demo]
  {Coupling Chirality, Polar Order, and Altermagnetic Spin Splitting in a Hybrid Manganese Chloride}

\abbreviations{IR,NMR,UV}

\begin{document}

\begin{tocentry}

  \includegraphics[width=\linewidth]{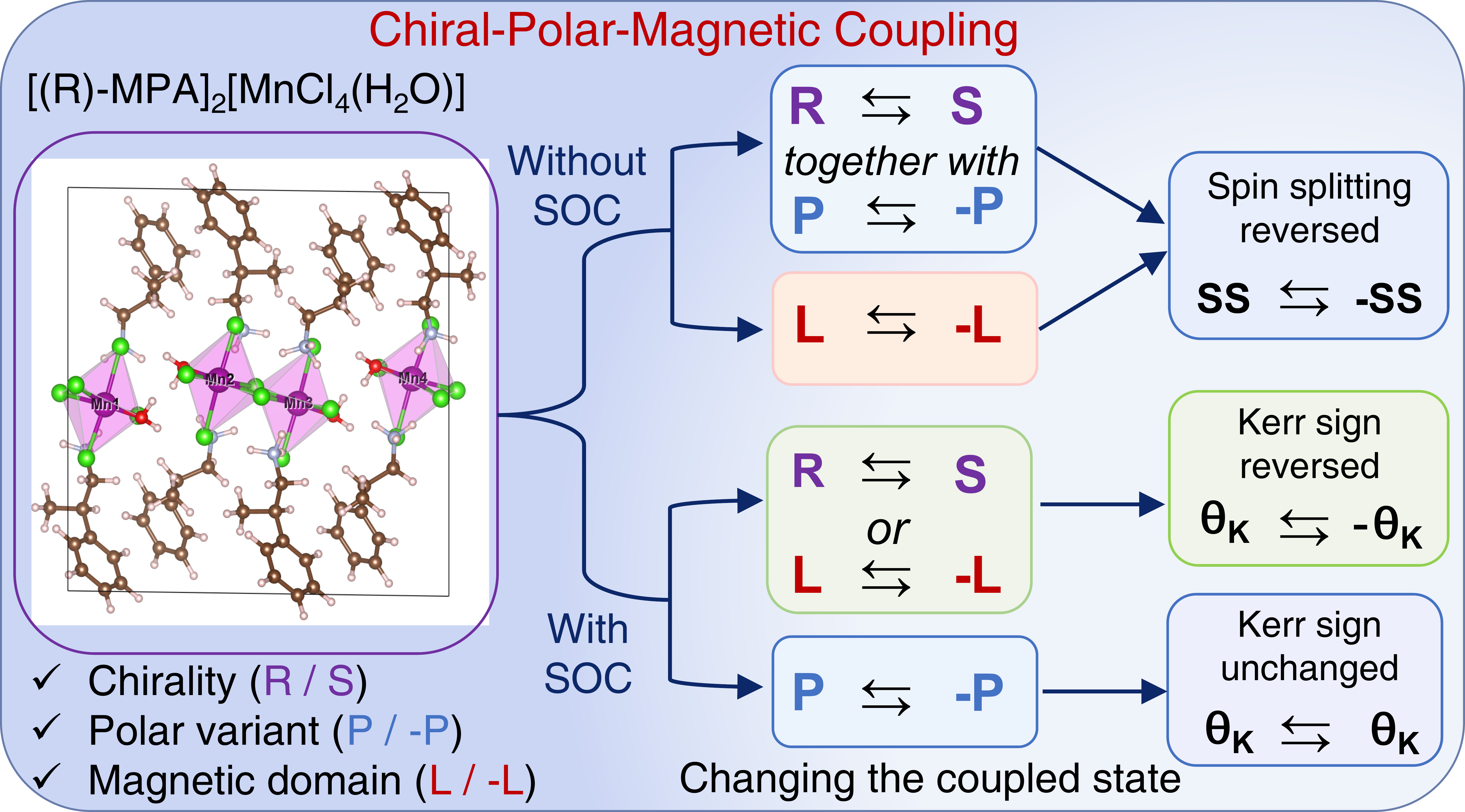}
 
\end{tocentry}

\begin{abstract}
Hybrid manganese halides enable the coexistence of molecular chirality, polar order, and magnetic exchange within a single lattice. Here, we combine first-principles calculations with spin-space-group analysis to investigate the synthesized enantiomeric pair $[(R)/(S)\text{-MPA}]_2[\text{MnCl}_4(\text{H}_2\text{O})]$ (MPA = $\beta$-methylphenethylammonium). We predict that its compensated magnetic state hosts altermagnetic spin splitting in the nonrelativistic limit, and that the coupled chiral, polar, and magnetic degrees of freedom define a symmetry-related manifold. From this manifold, we derive simple sign rules for the electronic and magneto-optical response: reversing both chirality and polarity, or reversing the magnetic domain alone, inverts the spin splitting throughout the Brillouin zone, whereas reversing chirality alone or polarity alone changes the spin-splitting sign only in symmetry-selected regions. With spin–orbit coupling, reversing chirality or magnetic order flips the Kerr rotation angle, while changing the polar variant leaves it unchanged. These results reveal a chemically accessible route to translate molecular handedness into symmetry-controlled spin splitting and magneto-optical readout in hybrid manganese halides. Critically, we show that the sign and momentum pattern of the splitting are governed by the interplay of the chiral, polar, and magnetic degrees of freedom. This interplay opens the possibility to control the spin splitting through a judicious design of the organic cations, by modulating their chirality and polarity.
\end{abstract}

\section{Introduction}
Altermagnets have been identified as a novel class of magnetic materials because they exhibit both zero net magnetization and spin-split electronic band structures\cite{AM,AM1,AM2,AM3}. In particular, the spin splitting arising from exchange interactions may be  much larger than that from spin-orbit coupling\cite{AM4,AM7}. Recently, many studies have proposed various altermagnetic material candidates\cite{RuO2,MnF2,B-AM2,B-AM3,B-AM4,Ding}. Specifically, time-reversal breaking altermagnets were predicted to host novel spin-related transport properties, including spin currents and the anomalous Hall effect, that were previously associated with ferromagnetism\cite{CrSb,Fe2O3,CuF2}. These unique properties substantially broaden the range of materials suitable for spintronics\cite{AM5,AM6}. So far, previously studied altermagnetic material candidates have mostly considered inorganic systems. While recent work has begun to explore hybrid and chiral altermagnetic candidates\cite{zhou,JACS,JACS2,Nanote}, the symmetry control of altermagnetic spin splitting by coupled molecular chirality, polar variants, and magnetic domains in chiral hybrid metal halides remains largely unexplored.

The possibility of  switchable spin splitting in  altermagnets controlled by an external field, such as an electric field, has become a critical challenge for device applications\cite{CASE1,CASE2,CASE3,CASE4,CASE5}. A notable example is a Cr-based metal-organic framework, where reversing the electric  polarization simultaneously switches the sign of spin splitting\cite{CASE2}. Magnetic hybrid organic--inorganic metal halides present another promising avenue for controlling spin splitting  due to their high structural tunability\cite{zhou}. Interestingly, they may introduce an additional degree of freedom arising from the chirality of the  organic cations\cite{HOIP1,HOIP2,HOIP3}.

Chiral organic–inorganic hybrid metal halides exhibit high structural tolerance, high compositional flexibility, and chemical tunability. Their organic cations provide the source of chirality and polarity, which can be transmitted throughout the entire crystal structure, while the magnetic ions within the inorganic framework give rise to magnetic ordering\cite{CME1,CME2,CME3,CME4}. On this basis, we find that the coupling among chirality, polarity, and altermagnetism can establish an interesting new  chiral–polar–magnetic materials platform. Access to symmetry-related coupled states through chirality, polarity, and magnetic-domain control enables predictable manipulation of the altermagnetic spin-splitting texture and related responses. In particular, global reversal of the spin splitting is achieved either by magnetic-domain reversal or by the combined reversal of chirality and polarity. Chiral polar organic–inorganic hybrid metal halides therefore emerge as a promising platform for symmetry-controlled altermagnetic states and functionalities.

This material is especially appealing as an experimentally realized test case. The $[(R)/(S)\text{-MPA}]_2[\text{MnCl}_4(\text{H}_2\text{O})]$ enantiomers have been synthesized and shown to exhibit chirality-dependent optical responses, while their low-temperature magnetic behavior has been interpreted as arising from a canted antiferromagnetic state with weak-ferromagnetic character\cite{MnCl4,2MnCl4}. These observations motivate a more microscopic symmetry analysis. In particular, they raise two central questions: whether the compensated magnetic backbone of this compound supports altermagnetic spin splitting, and how the corresponding spin texture and Kerr response transform under chirality reversal, polar-variant reversal, and magnetic-domain reversal. Here, we address these questions by combining first-principles calculations with spin-space-group symmetry. Rather than treating the compound only as a candidate material, we use it as a model platform to establish symmetry rules connecting molecular handedness, polar order, and magnetic order to electronic and magneto-optical observables.

Throughout this work, we keep these degrees of freedom clearly distinct. The labels $R$ and $S$ refer to the molecular handedness selected by synthesis and therefore identify different enantiomeric crystals. By contrast, $P$ and $-P$ denote symmetry-related polar variants of a given enantiomeric structure, while $+L$ and $-L$ denote opposite magnetic domains, corresponding to reversal of the Néel order. Thus, chirality selection, polar-variant reversal, and magnetic-domain reversal are treated as separate operations. Accordingly, the term ``symmetry-related coupled states'' is used to describe the full chiral–polar–magnetic configuration. Experimental access to each state will depend on the feasibility of enantiomer selection, polar-variant reversal, and magnetic-domain control in a given sample. Our calculations show that simultaneously reversing chirality and polarity, or reversing the magnetic-domain alone, leads to a reversal of spin splitting in the electronic band structure. Magneto-optical calculations further indicate that reversing either chirality or magnetic-domain changes the sign of the Kerr rotation angle, whereas reversing polar-variant alone does not alter the Kerr rotation angle.

\section{Results and Discussion}
\subsection{Crystal and Magnetic Configuration}
$[(R)/(S)\text{-MPA}]_2[\text{MnCl}_4(\text{H}_2\text{O})]$ is a one-dimensional chiral hybrid manganese chloride material recently synthesized by solvent-evaporation crystallization\cite{2MnCl4,MnCl4}. Figure~1a,b shows its crystal structure. Both enantiomers crystallize in the polar chiral space group P2$_1$. In the coordinate convention adopted here, the $y$ axis is chosen along the crystallographic $b$ axis, which is also the polar direction. The Cartesian axes used in the figures relate to the crystallographic axes as $x \parallel a$, $y \parallel b$, and $z$ nearly aligned with $c$ owing to the monoclinic angle $\beta = 90.78^\circ$. The structures of the $(R)$ and $(S)$ enantiomers are mirror images, related by the mirror operation $\mathcal{M}_y$ (reflection perpendicular to the $y$ axis). Their lattice constants are nearly identical ($a = 17.196$ \text{\AA}, $b = 7.187$ \text{\AA}, $c = 17.366$ \text{\AA}). 

Each Mn$^{2+}$ ion is octahedrally coordinated by five Cl ligands and one H$_2$O molecule. Several Cl atoms bridge adjacent Mn centers, so the local coordination number differs from the stoichiometric Cl/Mn ratio in the chemical formula.  The manganese chloride hydrate framework forms chain-like motifs running along $y/b$, separated by organic cations along $z$ (Figures~1c,d). By incorporating enantiopure $(R)$- or $(S)$-$\beta$-methylphenethylammonium (MPA$^+$) cations, the hybrid lattice acquires a definite handedness, loses all the improper symmetries, and inherits chiral information from the organic molecular sublattice.

%%%%%%%%%%%%%%%%%%%%%%%%%%%%%%%%%%%%%%%%%%%%%%%%%
\begin{figure}[htbp]
  \includegraphics[width=\linewidth]{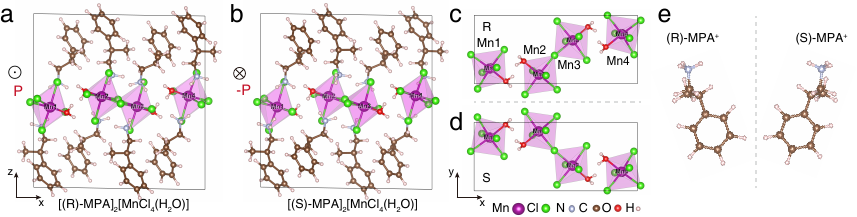}
  \caption{(a,b) Side views of the $(R)$- and $(S)$-enantiomers, respectively. The symbols $\odot$ and $\otimes$ denote the polarization direction, perpendicular to the paper plane outward along the $-y$ axis and inward along the $y$ axis, respectively. (c,d) Inorganic frameworks of the two enantiomers. (e) Opposite configurations of the MPA$^+$ stereogenic center. The ammonium-bearing side chain adopts opposite orientations relative to the phenyl-containing molecular framework in the two enantiomers. The gray dashed line denotes the mirror relationship.}
\end{figure}
%%%%%%%%%%%%%%%%%%%%%%%%%%%%%%%%%%%%%%%%%%%%%%%
The asymmetric unit contains four crystallographically independent $(S)/(R)$-MPA$^+$ cations. In each MPA$^+$ cation, chirality arises from the tetrahedral stereogenic carbon bonded to four different substituents: the phenyl group, the methyl group, the ammonium-bearing side chain, and hydrogen. The $R$ and $S$ forms correspond to opposite absolute configurations of this stereocenter. In Figure~1e, this molecular handedness is visualized by the opposite orientation of the ammonium-bearing side chain with respect to the phenyl-containing molecular framework.  Each MPA$^+$ cation possesses an intrinsic electric dipole moment arising from the asymmetric charge distribution between the ammonium moiety and the phenyl/alkyl groups. The dipole moment is oriented approximately along the axis connecting the phenyl carbon to the nitrogen. The out-of-plane components of the dipole moments along $z$ cancel each other, while the in-plane components remain, yielding a net polarization along the $-y/y$ directions (Figure~S1). Polarization magnitudes were evaluated with the Berry-phase method, giving nearly equal values of $0.232$ $\mu$C·cm$^{-2}$ along $-y$ and $y$. The Berry-phase polarization is used here to assign the symmetry-related polar branch; this work does not establish a ferroelectric switching pathway.

We next investigate the magnetic structure using $[(R)\text{-MPA}]_2[\text{MnCl}_4(\text{H}_2\text{O})]$ as a representative example. The structural/magnetic cell contains four Mn$^{2+}$ ions (Mn1–Mn4 in Figure~1c). Each Mn center carries a local spin magnetic moment of approximately $4.6~\mu_B$, consistent with a high-spin Mn(II) state with a $3d^5$ electronic configuration. The value is close to the ideal $S=5/2$ moment expected for Mn$^{2+}$, with the small reduction reflecting metal–ligand hybridization and the finite local integration volume used to define the Mn-projected moment.  These Mn sites form two symmetry-related pairs: Mn1/Mn4 and Mn2/Mn3. Within each pair, the ions are related by an intrinsic symmetry operation consisting of a twofold rotation about $y$ combined with a half-lattice translation along $y$. We introduce the magnetic order parameters $\boldsymbol{L}_{14} = \boldsymbol{S}_{\mathrm{Mn1}} - \boldsymbol{S}_{\mathrm{Mn4}}$ and $\boldsymbol{L}_{23} = \boldsymbol{S}_{\mathrm{Mn2}} - \boldsymbol{S}_{\mathrm{Mn3}}$. The spin orientation of each Mn$^{2+}$ ion can be up ($\uparrow$) or down ($\downarrow$). Considering the symmetry-relevant collinear configurations,  we obtain eight spin configurations (Table~1). Here, $0^{+}$/$0^{-}$ denotes parallel spin orientations ($\uparrow\uparrow$/$\downarrow\downarrow$) at the two magnetic sites, while $+$/$-$ denotes antiparallel ($\uparrow\downarrow$/$\downarrow\uparrow$). Configurations highlighted in white and gray are related by time-reversal $T$.

\begin{table}
  \caption{Relative energies $\Delta$E (meV/cell) of the eight magnetic configurations and their corresponding magnetic phases in $[(R)\text{-MPA}]_2[\text{MnCl}_4(\text{H}_2\text{O})]$. The energy of the configuration with $(\boldsymbol{L}_{14},\boldsymbol{L}_{23})=(+,-)$ is taken as the reference zero. ($\uparrow$) and ($\downarrow$) represent the spin orientation of Mn$^{2+}$ ions. Configurations highlighted in white and gray are related by $T$. AM, FM and CFiM denote altermagnetic, ferromagnetic and compensated ferrimagnetic states, respectively.}
      \rowcolors{2}{gray!15}{white}
  \begin{tabular}{lllllllll}
    \hline
    $\boldsymbol{L}_{14}$ & $\boldsymbol{L}_{23}$ & Mn1 & Mn2 & Mn3& Mn4& $\Delta$E (meV/cell) &Spin Space Group &Magnetism \\
    \hline
    $+$ & $+$ &$\uparrow$ &$\uparrow$ &$\downarrow$ &$\downarrow$ &0.7 &$P^{\overline{1}} 2_{1}^{\infty_{001} m} 1$& AM \\
    $-$ & $-$ &$\downarrow$ &$\downarrow$ &$\uparrow$ & $\uparrow$ &  0.7 &$P^{\overline{1}} 2_{1}^{\infty_{001} m} 1$& AM \\
    $+$ & $-$ &$\uparrow$ &$\downarrow$ &$\uparrow$ & $\downarrow$ & 0.0 &$P^{\overline{1}} 2_{1}^{\infty_{001} m} 1$& AM \\
    $-$ & $+$ & $\downarrow$ &$\uparrow$ &$\downarrow$ & $\uparrow$ & 0.0 &$P^{\overline{1}} 2_{1}^{\infty_{001} m} 1$& AM \\
    $0^{+}$ & $0^{+}$ &$\uparrow$ &$\uparrow$ &$\uparrow$ & $\uparrow$ &   45.5 &$P^{{1}} 2_{1}^{\infty_{001} m} 1$& FM \\
    $0^{-}$ & $0^{-}$ &$\downarrow$ &$\downarrow$ &$\downarrow$ & $\downarrow$ &  45.5 &$P^{{1}} 2_{1}^{\infty_{001} m} 1$& FM \\
    $0^{+}$ & $0^{-}$ &$\uparrow$ &$\downarrow$ &$\downarrow$ & $\uparrow$ &  44.2 &$P^{{1}} 2_{1}^{\infty_{001} m} 1$& CFiM \\
    $0^{-}$ & $0^{+}$ &$\downarrow$ &$\uparrow$ &$\uparrow$ & $\downarrow$ &   44.2 &$P^{{1}} 2_{1}^{\infty_{001} m} 1$ & CFiM \\
    \hline
  \end{tabular}
\end{table}

The different magnetic configurations are characterized by  distinct spin-space groups (SSGs)\cite{AM2,SSG1,SSG2}. Using the online program FINDSPINGROUP\cite{SSG3}, we identified the SSGs of the different configurations. The $(0^+,0^+)$ and $(0^-,0^-)$ states have SSG $P^{{1}} 2_{1}^{\infty_{001} m} 1$, corresponding to the ferromagnetic phase. For $(0^+,0^-)$ and $(0^-,0^+)$, the SSG is $P^{{1}} 2_{1}^{\infty_{001} m} 1$, which contains the symmetry $\{1 \parallel 2_{010}\}$. Owing to the absence of symmetry operations linking opposite spin sublattices, the system exhibits a nearly compensated ferrimagnetic state. In contrast, the $(+,+)$, $(-,-)$, $(+,-)$ and $(-,+)$ configurations belong to SSG $P^{\overline{1}} 2_{1}^{\infty_{001} m} 1$. Here, the operation $\{-1 \parallel 2_{010}\}$ relates magnetic sublattices with opposite spin orientations, identifying these four orders as altermagnetic phases.

Experimental studies report weak ferromagnetism below 3.2 K in $[(R)/(S)\text{-MPA}]_2[\text{MnCl}_4(\text{H}_2\text{O})]$, attributed to a canted antiferromagnetic state induced by Dzyaloshinskii–Moriya interactions\cite{MnCl4}. The lowest-energy state is the compensated altermagnetic $(\boldsymbol{L}_{14},\boldsymbol{L}_{23})=(+,-)$ (Table~1). In order to isolate the altermagnetic contribution to the Kerr response, we perform SOC-including MOKE calculations on the collinear compensated magnetic backbone, keeping the Néel-vector direction fixed. This reference description does not include weak SOC-induced canting, which, in a fully noncollinear treatment, could give rise to an additional small conventional magneto-optical signal. Thus, the sign rules analyzed below should be understood as symmetry rules for the altermagnetic Kerr component.

The above results show that chirality, polarity, and altermagnetic spin splitting can coexist in this compound. We construct different configurations  using these three degrees of freedom as order parameters and investigate the chiral--polar--magnetic coupling. To clearly illustrate coupling among chirality ($R/S$), polarity ($P/ -P$), and magnetic order ($\boldsymbol{L}_{14}, \boldsymbol{L}_{23}$), we represent the four associated order parameters as Ising-like doublets, each taking a value $+$ or $-$ to characterize the corresponding physical state. We restrict the coupled-state manifold to the altermagnetic ground-state doublet $(\boldsymbol{L}_{14}, \boldsymbol{L}_{23}) = (+,-)$ and $(-,+)$. (The $(+,+)$ and $(-,-)$ altermagnetic configurations lie only $\sim 0.7$ meV/cell higher in the reference PBE+U/DFT-D2 calculation, but the same energy ordering is retained under the $U_{\text{eff}}$ and vdW tests reported in Table~S2; thus the sign rules below are symmetry-dictated and do not rely on this small energy splitting.) This yields $2$ (chirality) $\times 2$ (polarity) $\times 2$ (magnetic domain) $= 8$ coupled states. Taking $[(R)\text{-MPA}]_2[\text{MnCl}_4(\text{H}_2\text{O})]$ as an example, the compound exhibits $R$ chirality, polarization along $-y$, and $(+,-)$ magnetic ordering, corresponding to the state $(+, +, +, -)$ for (chirality, polarization, $\boldsymbol{L}_{14}, \boldsymbol{L}_{23}$). In $[(S)\text{-MPA}]_2[\text{MnCl}_4(\text{H}_2\text{O})]$, chirality changes to $S$, polarization reverses to $+y$, and magnetic ordering remains unchanged, yielding $(-, -, +, -)$. The possible combinations give eight coupled-state labels: $(+, +, +,- )$, $(+, -, +, -)$, $(-, +, +, -)$, $(-, -, +, -)$, $(+, +, -, +)$, $(+, -, -, +)$, $(-, +, -, +)$, $(-, -, -, +)$. Crystal symmetry operations transform these states into each other. A mirror operation parallel to the dipole axis reverses chirality while preserving polarity; a twofold rotation perpendicular to the dipole axis reverses polarity while preserving chirality. Time reversal flips the magnetic-domain. Within the spin space group formalism, the connecting symmetry operations include $\{1 \parallel 2_{001}\}$, $\{1 \parallel M_{100}\}$, $\{1 \parallel M_{010}\}$, and $\{-1 \parallel 1\}$. Figure~S2 schematically illustrates these relations. Here, $\{1 \parallel 2_{001}\}$ is a twofold rotation about $z$, $\{1 \parallel M_{100}\}$ and $\{1 \parallel M_{010}\}$ are mirror reflections across the $yz$ and $xz$ planes, respectively, and $\{-1 \parallel 1\}$ is time reversal. $\{1 \parallel 2_{001}\}$, $\{1 \parallel M_{100}\}$, $\{1 \parallel M_{010}\}$ are used as domain operations that map between degenerate ferroic/enantiomeric states in the absence of external fields; they are not necessarily elements of the space group of a single domain.

\subsection{Symmetry Analysis and Spin-Degenerate Manifolds}

 We take the $(+,+,+,-)$ state as a representative example to analyze the spin symmetry of the electronic bands. The energy bands as a function of spin ($s$) and crystal momentum ($\boldsymbol{k}$) are written as $E(s,\boldsymbol{k})$. Spin degeneracy means $E(s,\boldsymbol{k}) = E(-s,\boldsymbol{k})$, whereas altermagnetic splitting means $E(s,\boldsymbol{k}) \neq E(-s,\boldsymbol{k})$, with opposite-spin states related at a symmetry-connected momentum $\boldsymbol{k}'$: $E(s,\boldsymbol{k}) = E(-s,\boldsymbol{k}')$. The $(+,+,+,-)$ state contains the spin space group operations $\{1||1\}$ and $\{-1||2_{010}\}$. Applying $\{-1||2_{010}\}$ to $E(s,\boldsymbol{k})$ yields $E(s,k_x,k_y,k_z) = E(-s,-k_x,k_y,-k_z)$. For $k_x = 0$ and $k_z = 0$, this simplifies to $E(s,k_y) = E(-s,k_y)$, indicating that the bands are spin-degenerate along the $\Gamma(0,0,0)$--$Z(0,0.5,0)$ path. In the nonrelativistic collinear limit, each spin-channel Hamiltonian can be chosen real, giving $E(s,\boldsymbol{k}) = E(s,-\boldsymbol{k})$, from which one obtains $E(s,k_x,k_y,k_z) = E(-s,k_x,-k_y,k_z)$. This relation shows that spin degeneracy also occurs in the $k_y = 0$ and $k_y = \pm 0.5$ planes in fractional reciprocal coordinates (i.e., $k_y = 0$ and $k_y = \pm \pi/b$ in Cartesian reciprocal units). In the Brillouin zone schematic of Figure~2a, the red points, lines, and planes mark the regions of spin degeneracy. In bulk $d$-wave altermagnets, two spin-degenerate nodal surfaces cross the $\Gamma$ point\cite{AM2,x-wave}. By contrast, in the present $(+,+,+,-)$ state, the absence of the corresponding mirror symmetry allows only one such surface to cross $\Gamma$ (Figure~2a). Thus, the bulk compound has lower symmetry than a $d$-wave altermagnet.

%%%%%%%%%%%%%%%%%%%%%%%%%%%%%%%%%%%
\begin{figure}[htbp]
  \includegraphics[width=\linewidth]{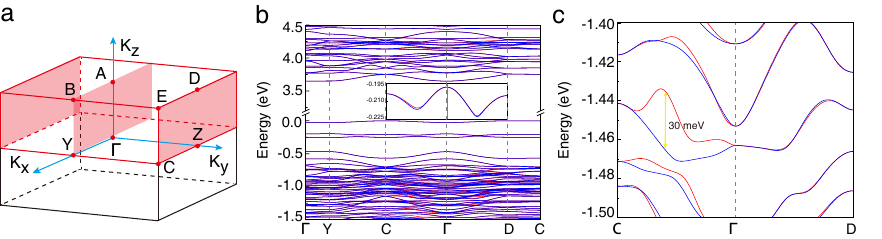}
  \caption{(a) Schematic of the first Brillouin zone for the $(+,+,+,-)$ configuration, with red points, lines, and planes denoting the regions of spin degeneracy. (b, c) Nonrelativistic electronic band structures of the $(+,+,+,-)$ state along the high-symmetry path, with the Fermi level set as the zero of energy. The inset in (b) and panel (c) highlight representative spin-splitting features along the C--$\Gamma$--D path in the valence-band region near $E=-1.46$ eV.}
\end{figure}
%%%%%%%%%%%%%%%%%%%%%%%%%%%%%%%%%%%

The electronic band structure of the $(+,+,+,-)$ state (Figure~2b) exhibits an insulating gap of approximately $3.65~\mathrm{eV}$. Along the symmetry-protected $\Gamma$(0,0,0)--Y(0.5,0,0)--C(0.5,0.5,0) and D(0,0.5,0.5)--C(0.5,0.5,0) paths, the bands remain spin-degenerate. In contrast, pronounced spin splitting appears along the C--$\Gamma$--D path, in full agreement with the symmetry analysis. Figure~2c provides a magnified view of the bands along this path within the energy window from $-1.5$ to $-1.4~\mathrm{eV}$, where the maximum spin splitting reaches $30~\mathrm{meV}$.

\subsection{Coupled State Labeling and Sign Rules for Spin Splitting}
As discussed above, these energy-equivalent states are related by symmetry operations.
To analyze the relation between the spin splittings of different coupled states, we take the $(+,+,+,-)$ state as a reference. The $(+,+,+,-)$ state and the other coupled states are connected by crystal symmetry operations and therefore exhibit symmetry-related spin splittings. Without spin-orbit coupling, the constraints imposed by these operations allow us to derive a correspondence between the spin-splitting signatures of the other states and that of the $(+,+,+,-)$ state.

%%%%%%%%%%%%%%%%%%%%%%%%%%%%%%%%%%%
\begin{figure}[htbp]
  \includegraphics[height=10cm,keepaspectratio]{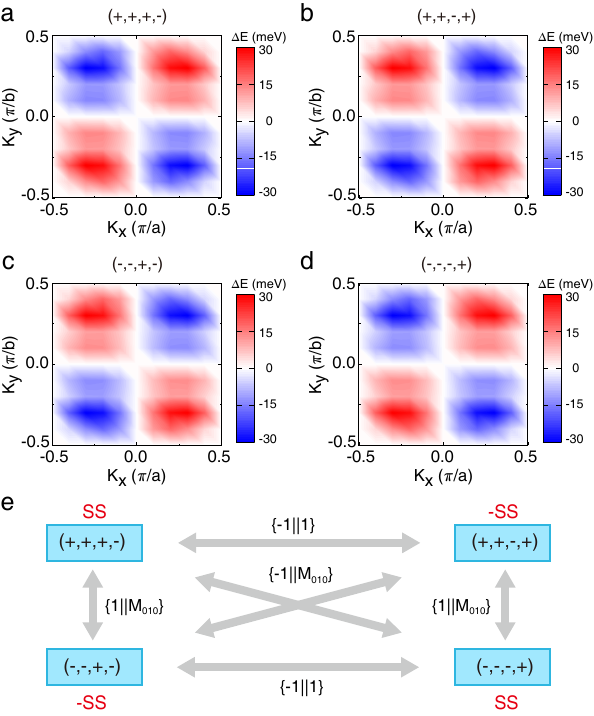} % 固定高度4cm，保持比例
  \caption{(a-d) Spin splitting projection of the coupled states on the $k_z = 0$ plane near $E = -1.46$ eV. (e) Schematic illustration of the spin splitting signs of coupled states connected by different symmetry operations, with the $(+,+,+,-)$ state as the reference.}
\end{figure}
%%%%%%%%%%%%%%%%%%%%%%%%%%%%%%%%%%%

For instance, in the $(-,-,+,-)$ state, the intrinsic $\{-1||2_{010}\}$ operation of the system combined with the $\{1||M_{010}\}$ operation that connects to the $(+,+,+,-)$ state is equivalent to $\{-1||-1\}$, which reverses the sign of the spin splitting throughout the entire Brillouin zone. A detailed symmetry analysis and derivation are provided in the Supporting Information (SI). Based on this analysis, Figure~S3 schematically illustrates the spin-splitting features in the $k_x = 0$ and $k_z = 0$ planes. The results clearly show that when chirality and polarity are reversed together, or when the magnetic moment direction is changed, the sign of the spin splitting is fully reversed. In contrast, reversing chirality or polarity alone changes the sign only in selected  regions in reciprocal space.

To explicitly verify the symmetry-derived relations, we computed the spin-splitting projections onto the $k_z = 0$ plane for the pair of bands near $E = -1.46$~eV in four representative coupled states: $(+,+,+,-)$, $(-,-,+,-)$, $(+,+,-,+)$, and $(-,-,-,+)$ (Figures~3a--d). For completeness, the band structures of all eight coupled states along the high-symmetry C--$\Gamma$--D path are reported in Figure~S4. These calculations confirm the symmetry analysis. Taking the $(+,+,+,-)$ configuration as the reference state, we denote its spin-splitting sign as $\mathrm{SS}$, while $-\mathrm{SS}$ indicates the reversed sign. Figure~3e summarizes the symmetry-imposed sign relations among the four representative states, whereas Figure~S5 extends the same analysis to the remaining four coupled states.

In the hydrated manganese chloride hybrid studied here, crystal symmetry imposes a well-defined correspondence between molecular chirality, polar variant, magnetic domain, and the resulting spin-splitting texture. Table~2 summarizes the spin-splitting signatures of the eight symmetry-related coupled states. Within this chiral--polar--magnetic framework, the spin-splitting pattern is therefore not arbitrary, but is fixed by the selected coupled state. In particular, magnetic-domain reversal globally reverses the spin-splitting texture, while the combined reversal of chirality and polarity produces the same global sign change. By contrast, reversing chirality or polarity alone modifies the texture according to the corresponding symmetry operation, without simply inverting it everywhere in momentum space. These results provide a symmetry-based microscopic picture of the coupling among chirality, polar order, and magnetic order, and suggest a design principle for hybrid magnetic materials with controllable spin-dependent optoelectronic responses.

\subsection{Magneto-optical Kerr effect}
The magneto-optical Kerr effect (MOKE) originates from the off-diagonal components of the dielectric tensor induced by spin-orbit coupling and serves as a sensitive probe of electronic structure, exchange interactions, and symmetry breaking in magnetic systems. The sign and orientation of the MOKE signal are strictly constrained by magnetic symmetry. We first determined the magnetocrystalline anisotropy of the $(+,+,+,-)$ state by comparing configurations in which the Néel vector is oriented along the $x$, $y$, and $z$ directions. The results identify the $xy$ plane as the easy plane, with an anisotropy energy of about $0.1$~meV/cell, indicating a weak preference for in-plane Néel-vector orientation. In the SOC-including MOKE calculations, we therefore fixed the Néel vector along the $x$ axis as a representative in-plane direction. Under this choice, the magnetic symmetry is described by the magnetic space group $P2_1$. The corresponding symmetry-allowed form of the dielectric tensor was obtained using the \texttt{MTENSOR} program of the Bilbao Crystallographic Server and is given by\cite{MOKE}:

\begin{equation}
\varepsilon=\left[\begin{array}{ccc}
\varepsilon_{x x} & 0 & \varepsilon_{x z} \\
0 & \varepsilon_{y y} & 0 \\
\varepsilon_{z x} & 0 & \varepsilon_{z z}
\end{array}\right] 
\end{equation}
Because the dielectric tensor relates to the optical conductivity through
$\varepsilon_{i j}(\omega)= \delta_{i j}+i \frac{4 \pi}{\omega} \sigma_{i j}(\omega)$\cite{MOKE3,MOKE4}, the optical conductivity in the $(+,+,+,-)$ state can be written as:
\begin{equation}
\sigma=\left[\begin{array}{ccc}
\sigma_{x x} & 0 & \sigma_{x z} \\
0 & \sigma_{y y} & 0 \\
\sigma_{z x} & 0 & \sigma_{z z}
\end{array}\right] 
\end{equation}
%The magnetization direction of the $(+,+,+,+)$ state lies in-plane, such that its MOKE belongs to the longitudinal-MOKE. 
The  Kerr rotation angle $\theta_K$ and ellipticity $\eta_K$ are  combined into the complex Kerr angle, given by\cite{MOKE1,MOKE2}:

\begin{equation}
\phi_{K}=\theta_{K}+i \eta_{K}=\frac{-\left(\sigma_{x z}-\sigma_{z x}\right)}{2 \sigma_{0} \sqrt{1+\frac{i 4 \pi}{\omega} \sigma_{0}}}
\end{equation}
 where $ \sigma_{0}=\left(\sigma_{xx}+\sigma_{zz}\right) / 2$.
As indicated by Eq.~(3), the sign of $(\sigma_{xz}-\sigma_{zx})$ in the complex Kerr angle determines the direction of $\theta_{K}$. Under a magnetic symmetry operation $R$, the optical conductivity tensor transforms as
\begin{equation}
R {\sigma} R^{-1}={\sigma}^{\prime}
\end{equation}

In the presence of SOC, the eight coupled states are related by magnetic symmetry operations, which simultaneously transform spatial coordinates and magnetic-moment orientations, replacing the spin-space-group description used in the nonrelativistic limit. As an illustrative example, the magnetic symmetry operation $2_{001}$ transforms $(x,y,z)$ into $(-x,-y,z)$, and the magnetic moment $(m_x,m_y,m_z)$ into $(-m_x,-m_y,m_z)$. Applying $2_{001}$ maps the $(+,+,+,-)$ state onto the $(+,-,-,+)$ state. Subsequent time-reversal ($T$) of this $2_{001}$-generated state gives the $(+,-,+,-)$ state. Figure~4a summarizes the magnetic symmetry operations that connect the different coupled states. The explicit forms of the optical conductivity tensor constrained by these magnetic symmetries are provided in the SI. The symmetry analysis shows that $T$, $M_{100}$, and $2_{001}$ reverse the sign of $(\sigma_{xz}-\sigma_{zx})$ in Eq.~(3), whereas $M_{010}$ leaves it unchanged. Based on these considerations and taking the Kerr rotation angle of the $(+,+,+,-)$ state as a reference, the signs of the Kerr angles for the different coupled states are indicated in Figure~4a. The Kerr angle changes sign when either chirality or the magnetic domain is reversed, but not when only the polar variant is inverted.  Figures~4c,d show $\theta_K$ of the eight coupled states as a function of incident photon energy from DFT calculations. The relative signs among different states are fully consistent with the symmetry analysis. A sizable Kerr rotation is predicted near 5.6 eV within the present PBE+U+SOC framework. The corresponding Kerr ellipticity spectra are shown in Figure~S6. The absolute peak positions and amplitudes should be interpreted at the independent-particle PBE+U+SOC level. Excitonic effects, local-field effects, and denser optical sampling may shift the spectral features. The relative sign relations among coupled states, however, are dictated by symmetry and are therefore robust.
%%%%%%%%%%%%%%%%%%%%%%%%%%%%%%%%%%%
\begin{figure}[htbp]
  \includegraphics[width=\linewidth]{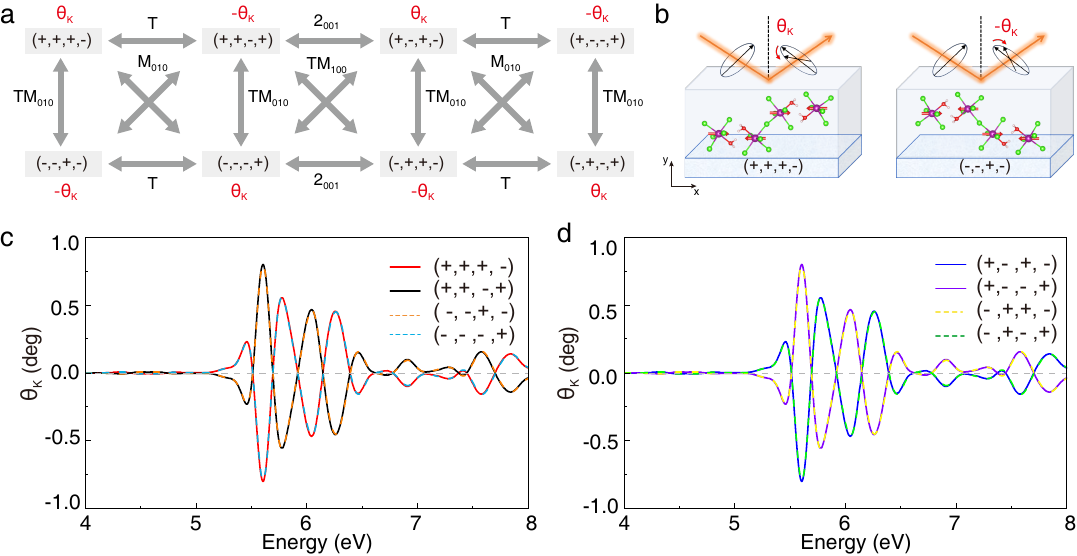}
  \caption{ (a) Schematic illustration of the Kerr angle signs for the eight coupled states connected by different symmetry operations, with the $(+,+,+,-)$ state as the reference. (b) Schematic of the magneto-optical Kerr effect, where the $(+,+,+,-)$ and $(-,-,+,-)$ states exhibit oppositely directed Kerr rotation angles. (c, d) Evolution of the Kerr angle as a function of incident photon energy for the eight coupled states. }
\end{figure}
%%%%%%%%%%%%%%%%%%%%%%%%%%%%%%%%%%%

Equation~(3) describes the Kerr response in the $x$--$z$ polarization plane for reflection from a $y$-normal surface, within the small-off-diagonal-conductivity approximation. Experimental comparison requires a fixed laboratory convention, the same surface orientation, and a selected magnetic domain. Domain averaging, surface selection, or weak SOC-induced canting may affect the amplitude or add a conventional magneto-optical background, but not the symmetry-imposed sign rule for the single-domain altermagnetic contribution. Figure~4b illustrates a schematic magneto-optical readout geometry based on the $(+,+,+,-)$ and $(-,-,+,-)$ states. These two coupled states correspond to the experimentally synthesized $[(R)\text{-MPA}]_2[\text{MnCl}_4(\text{H}_2\text{O})]$ and $[(S)\text{-MPA}]_2[\text{MnCl}_4(\text{H}_2\text{O})]$ materials, respectively, and exhibit opposite $\theta_{K}$ signs. Reversing the magnetic domain transforms the system from $(+,+,+,-)$/$(-,-,+,-)$ into $(+,+,-,+)$/$(-,-,-,+)$, thereby reversing the Kerr rotation direction. This switching rule also applies to the remaining four coupled states. Table~2 summarizes the Kerr angle signs of the eight coupled states. These results indicate that $[(R)/(S)\text{-MPA}]_2[\text{MnCl}_4(\text{H}_2\text{O})]$ and related systems provide a clear physical picture of the magneto-optical response in multi-degree-of-freedom systems and may serve as useful platforms for experimentally testing symmetry-controlled MOKE sign relations.

%%%%%%%%%%%%%%%%%%%%%%%%%%%%%%%%%%%%%%%%%%%%%%%%
\begin{table}
%\setcounter{table}{1}
%\renewcommand{\thetable}{S\arabic{table}}
%\centering
\caption{Summary of the spin splitting and Kerr angle signs for the eight coupled states. SS and SS' denote distinct momentum-dependent spin-splitting patterns, while $-$SS and $-$SS' denote the corresponding patterns with globally reversed spin polarization. The Kerr angle signs are defined with the $(+,+,+,-)$ state taken as the reference.}
\label{tab:S2}
\renewcommand{\arraystretch}{1.2} % 增加行间距（可根据需要调整数值）
\begin{tabular}{c|c|c}
\hline
States & Spin splitting sign& Kerr angle sign \\
\hline
$(+,+,+,-)$ & SS &  $\theta_K$  \\
\hline
$(-,-,+,-)$ & -SS & -$\theta_K$  \\
\hline
$(+,+,-,+)$ & -SS &  -$\theta_K$  \\
\hline
$(-,-,-,+)$ & SS &  $\theta_K$  \\
\hline
$(+,-,+,-)$ & SS' &  $\theta_K$  \\
\hline
$(-,+,+,-)$ & -SS' &  -$\theta_K$  \\
\hline
$(+,-,-,+)$ & -SS' &  -$\theta_K$  \\
\hline
$(-,+,-,+)$ & SS' &  $\theta_K$  \\
\hline
\end{tabular}
\end{table}

%%%%%%%%%%%%%%%%%%%%%%%%%%%%%%%%%%%%%%%%%%%%%%%%

In this work, we investigate $[(R)/(S)\text{-MPA}]_2[\text{MnCl}_4(\text{H}_2\text{O})]$ as a chiral polar magnetic hybrid in which molecular handedness, polar order, and altermagnetic spin splitting coexist within a single material platform. Microscopically, the enantiopure MPA$^+$ cations impose a handed, noncentrosymmetric distortion on the inorganic Mn--Cl--H$_2$O framework. This structural imprint couples the organic molecular sublattice to the compensated magnetic backbone, linking chirality, polarity, and magnetic order. More broadly, this study shows that chiral hybrid manganese halides offer a chemically versatile platform for symmetry-engineered altermagnetism. The mechanism identified here does not rely on a single molecular choice or a unique inorganic framework, but on general and tunable design elements: enantiopure organic cations, directional hydrogen-bonding networks, and magnetically active metal--halide units. The Mn--halide coordination motif can be extended from chlorides to bromide- and iodide-based manganese compounds, as well as to related transition-metal halide frameworks\cite{D-case1,D-case2,D-case3,D-case4}. At the same time, MPA$^+$ can be replaced by other chiral cations, such as chiral phenethylammonium derivatives, chiral amino alcohols, or substituted chiral alkylammonium ions\cite{D-case5,D-case6,D-case7}. Through their configuration, substituents, and hydrogen-bonding geometry, these organic building blocks can impose noncentrosymmetric distortions, transfer molecular handedness to the inorganic lattice, and tune the coupling to the compensated magnetic backbone. This combination of chemical modularity and symmetry control points to a broader family of chiral polar hybrids in which altermagnetic spin splitting and magneto-optical responses can be deliberately designed.

\section{Conclusions}
In conclusion, $[(R)/(S)\text{-MPA}]_2[\text{MnCl}_4(\text{H}_2\text{O})]$ reveals how molecular chirality, polar structure, and compensated magnetic order can be locked together by symmetry to control altermagnetic and magneto-optical responses. The selected chiral--polar--magnetic state fixes the altermagnetic spin-splitting texture, whose global sign is reversed either by switching the magnetic domain or by simultaneously reversing chirality and polar variant. By contrast, the Kerr response follows a different symmetry rule: $\theta_K$ changes sign under chirality or magnetic-domain reversal, but not under polar-variant reversal alone. These results establish a direct route for translating chemically encoded handedness into electronic spin textures and Kerr-active optical signatures. They further identify chiral hybrid metal halides as a versatile platform for designing compensated magnetic materials in which spin splitting and magneto-optical functionality are programmed by symmetry.

\begin{suppinfo}

Computational details; symmetry relations among coupled chiral–polar–magnetic states; robustness tests with different ($U_{\mathrm{eff}}$) values and van der Waals corrections; derivations of spin-splitting and optical-conductivity sign rules; additional dipole, band-structure, spin-splitting, Kerr-ellipticity, and structural figures (PDF).
\section{Data Availability Statement}
 The data underlying this study are available in the published article and its Supporting Information. Optimized structures and representative input files are available from the corresponding authors upon reasonable request.
 \section*{Notes}
 The authors declare no competing financial interest.
\end{suppinfo}

%%%%%%%%%%%%%%%%%%%%%%%%%%%%%%%%%%
\section{Acknowledgements}
This work is supported by the National Key Research and Development Plan Project of China (Grant No. 2025YFA1411100) and the Fundamental Research Funds for the Central Universities (Grant No. 2242025K30023). We also acknowledge support from the Open Research Fund of the Key Laboratory of Quantum Materials and Devices (Southeast University) and the Big Data Computing Center of Southeast University.
We acknowledge the support from the Italian Ministry of Research under the PRIN 2022 Grant No 2022F2K7J5 with the title “Two-dimensional chiral hybrid organic--inorganic perovskites for chiroptoelectronics” PE 3 funded by PNRR Mission 4 Istruzione e Ricerca - Component C2 - Investimento 1.1, Fondo per il Programma Nazionale di Ricerca e Progetti di Rilevante Interesse Nazionale PRIN 2022 – CUP B53D23004130006.
%%%%%%%%%%%%%%%%%%%%
\bibliography{achemso-demo}

@article{AM,
  author = {Zhu, Yu-Peng and Chen, Xiaobing and Liu, Xiang-Rui and Liu, Yuntian and Liu, Pengfei and Zha, Heming and Qu, Gexing and Hong, Caiyun and Li, Jiayu and Jiang, Zhicheng and Ma, Xiao-Ming and Hao, Yu-Jie and Zhu, Ming-Yuan and Liu, Wenjing and Zeng, Meng and Jayaram, Sreehari and Lenger, Malik and Ding, Jianyang and Mo, Shu and Tanaka, Kiyohisa and Arita, Masashi and Liu, Zhengtai and Ye, Mao and Shen, Dawei and Wrachtrup, J{\"o}rg and Huang, Yaobo and He, Rui-Hua and Qiao, Shan and Liu, Qihang and Liu, Chang},
  title = {Observation of plaid-like spin splitting in a noncoplanar antiferromagnet},
  journal = {Nature},
  year = {2024},
  volume = {626},
  number = {7999},
  pages = {523--528},
  doi = {10.1038/s41586-024-07023-w},
  url = {https://doi.org/10.1038/s41586-024-07023-w},
  issn = {1476-4687},
}

@article{AM1,
  title = {Emerging Research Landscape of Altermagnetism},
  author = {{\v S}mejkal, Libor and Sinova, Jairo and Jungwirth, Tomas},
  journal = {Phys. Rev. X},
  volume = {12},
  issue = {4},
  pages = {040501},
  numpages = {27},
  year = {2022},
  month = {Dec},
  publisher = {American Physical Society},
  doi = {10.1103/PhysRevX.12.040501},
  url = {https://link.aps.org/doi/10.1103/PhysRevX.12.040501}
}

@article{AM2,
  title = {Beyond Conventional Ferromagnetism and Antiferromagnetism: A Phase with Nonrelativistic Spin and Crystal Rotation Symmetry},
  author = {{\v S}mejkal, Libor and Sinova, Jairo and Jungwirth, Tomas},
  journal = {Phys. Rev. X},
  volume = {12},
  issue = {3},
  pages = {031042},
  numpages = {16},
  year = {2022},
  month = {Sep},
  publisher = {American Physical Society},
  doi = {10.1103/PhysRevX.12.031042},
  url = {https://link.aps.org/doi/10.1103/PhysRevX.12.031042}
}

@article{AM3,
  title = {Editorial: Altermagnetism---A New Punch Line of Fundamental Magnetism},
  author = {Mazin, Igor},
  collaboration = {The PRX Editors},
  journal = {Phys. Rev. X},
  volume = {12},
  issue = {4},
  pages = {040002},
  numpages = {3},
  year = {2022},
  month = {Dec},
  publisher = {American Physical Society},
  doi = {10.1103/PhysRevX.12.040002},
  url = {https://link.aps.org/doi/10.1103/PhysRevX.12.040002}
}

@article{AM4,
  title = {Large Band Splitting in $g$-Wave Altermagnet CrSb},
  author = {Ding, Jianyang and Jiang, Zhicheng and Chen, Xiuhua and Tao, Zicheng and Liu, Zhengtai and Li, Tongrui and Liu, Jishan and Sun, Jianping and Cheng, Jinguang and Liu, Jiayu and Yang, Yichen and Zhang, Runfeng and Deng, Liwei and Jing, Wenchuan and Huang, Yu and Shi, Yuming and Ye, Mao and Qiao, Shan and Wang, Yilin and Guo, Yanfeng and Feng, Donglai and Shen, Dawei},
  journal = {Phys. Rev. Lett.},
  volume = {133},
  issue = {20},
  pages = {206401},
  numpages = {7},
  year = {2024},
  month = {Nov},
  publisher = {American Physical Society},
  doi = {10.1103/PhysRevLett.133.206401},
  url = {https://link.aps.org/doi/10.1103/PhysRevLett.133.206401}
}

@article{AM5,
author = {Bai, Ling and Feng, Wanxiang and Liu, Siyuan and Šmejkal, Libor and Mokrousov, Yuriy and Yao, Yugui},
title = {Altermagnetism: Exploring New Frontiers in Magnetism and Spintronics},
journal = {Adv. Funct. Mater.},
volume = {34},
number = {49},
pages = {2409327},
doi = {https://doi.org/10.1002/adfm.202409327},
url = {https://advanced.onlinelibrary.wiley.com/doi/abs/10.1002/adfm.202409327},
eprint = {https://advanced.onlinelibrary.wiley.com/doi/pdf/10.1002/adfm.202409327},
year = {2024}
}

@article{AM7,
title = {Spin-split collinear antiferromagnets: A large-scale ab-initio study},
journal = {Mater. Today Phys.},
volume = {32},
pages = {100991},
year = {2023},
issn = {2542-5293},
doi = {https://doi.org/10.1016/j.mtphys.2023.100991},
url = {https://www.sciencedirect.com/science/article/pii/S2542529323000275},
author = {Yaqian Guo and Hui Liu and Oleg Janson and Ion Cosma Fulga and Jeroen {van den Brink} and Jorge I. Facio},
}

@article{RuO2,
author = {Fedchenko, Olena and Minár, Jan and Akashdeep, Akashdeep and D’Souza, Sunil Wilfred and Vasilyev, Dmitry and Tkach, Olena and Odenbreit, Lukas and Nguyen, Quynh and Kutnyakhov, Dmytro and Wind, Nils and Wenthaus, Lukas and Scholz, Markus and Rossnagel, Kai and Hoesch, Moritz and Aeschlimann, Martin and Stadtmüller, Benjamin and Kläui, Mathias and Schönhense, Gerd and Jungwirth, Tomas and Hellenes, Anna Birk and Jakob, Gerhard and Šmejkal, Libor and Sinova, Jairo and Elmers, Hans-Joachim},
title = {Observation of time-reversal symmetry breaking in the band structure of altermagnetic RuO$_2$},
journal = {Sci. Adv.},
volume = {10},
number = {5},
pages = {eadj4883},
year = {2024},
doi = {10.1126/sciadv.adj4883},
URL = {https://www.science.org/doi/abs/10.1126/sciadv.adj4883},
eprint = {https://www.science.org/doi/pdf/10.1126/sciadv.adj4883},
}

@article{MnF2,
  title = {Absence of Altermagnetic Magnon Band Splitting in ${\mathrm{MnF}}_{2}$},
  author = {Morano, V. C. and Maesen, Z. and Nikitin, S. E. and Lass, J. and Mazzone, D. G. and Zaharko, O.},
  journal = {Phys. Rev. Lett.},
  volume = {134},
  issue = {22},
  pages = {226702},
  numpages = {6},
  year = {2025},
  month = {Jun},
  publisher = {American Physical Society},
  doi = {10.1103/PhysRevLett.134.226702},
  url = {https://link.aps.org/doi/10.1103/PhysRevLett.134.226702}
}

@article{B-AM2,
  title = {General Stacking Theory for Altermagnetism in Bilayer Systems},
  author = {Pan, Baoru and Zhou, Pan and Lyu, Pengbo and Xiao, Huaping and Yang, Xuejuan and Sun, Lizhong},
  journal = {Phys. Rev. Lett.},
  volume = {133},
  issue = {16},
  pages = {166701},
  numpages = {7},
  year = {2024},
  month = {Oct},
  publisher = {American Physical Society},
  doi = {10.1103/PhysRevLett.133.166701},
  url = {https://link.aps.org/doi/10.1103/PhysRevLett.133.166701}
}

@article{B-AM3,
  title = {Stacking-dependent ferroicity of a reversed bilayer: Altermagnetism or ferroelectricity},
  author = {Sun, Wencong and Ye, Haoshen and Liang, Li and Ding, Ning and Dong, Shuai and Wang, Shan-Shan},
  journal = {Phys. Rev. B},
  volume = {110},
  issue = {22},
  pages = {224418},
  numpages = {7},
  year = {2024},
  month = {Dec},
  publisher = {American Physical Society},
  doi = {10.1103/PhysRevB.110.224418},
  url = {https://link.aps.org/doi/10.1103/PhysRevB.110.224418}
}

@article{B-AM4,
  title={Ferrovalley Physics in Stacked Bilayer Altermagnetic Systems},
  author={Li, Yun-Qin and Zhang, Yu-Ke and Lu, Xin-Le and Shao, Ya-Ping and Bao, Zhi-Qiang and Zheng, Jun-Ding and Tong, Wen-Yi and Duan, Chun-Gang},
  journal={Nano Lett.},
  volume={25},
  number={15},
  pages={6032--6039},
  year={2025},
  publisher={ACS Publications}
}

@article{Ding,
  title = {Ferroelastically tunable altermagnets},
  author = {Ding, Ning and Ye, Haoshen and Wang, Shan-Shan and Dong, Shuai},
  journal = {Phys. Rev. B},
  volume = {112},
  issue = {22},
  pages = {L220410},
  numpages = {8},
  year = {2025},
  month = {Dec},
  publisher = {American Physical Society},
  doi = {10.1103/m33v-xwn3},
  url = {https://link.aps.org/doi/10.1103/m33v-xwn3}
}

@article{CrSb,
  title={N{\'e}el vector-dependent anomalous transport in altermagnetic metal CrSb},
  author={Yu, Tianye and Shahid, Ijaz and Liu, Peitao and Shao, Ding-Fu and Chen, Xing-Qiu and Sun, Yan},
  journal={npj Quantum Mater.},
  volume={10},
  number={1},
  pages={47},
  year={2025},
  publisher={Nature Publishing Group UK London}
}

@article{Fe2O3,
author = {Galindez-Ruales, Edgar and Gonzalez-Hernandez, Rafael and Schmitt, Christin and Das, Shubhankar and Fuhrmann, Felix and Ross, Andrew and Golias, Evangelos and Akashdeep, Akashdeep and Lünenbürger, Laura and Baek, Eunchong and Yang, Wanting and Šmejkal, Libor and Krishna, Venkata and Jaeschke-Ubiergo, Rodrigo and Sinova, Jairo and Rothschild, Avner and You, Chun-Yeol and Jakob, Gerhard and Kläui, Mathias},
title = {Revealing the Altermagnetism in Hematite via XMCD Imaging and Anomalous Hall Electrical Transport},
journal = {Adv. Mater.},
volume = {37},
number = {41},
pages = {e05019},
doi = {https://doi.org/10.1002/adma.202505019},
url = {https://advanced.onlinelibrary.wiley.com/doi/abs/10.1002/adma.202505019},
eprint = {https://advanced.onlinelibrary.wiley.com/doi/pdf/10.1002/adma.202505019},
year = {2025}
}

@article{CuF2,
doi = {10.1088/0256-307X/42/8/080705},
url = {https://doi.org/10.1088/0256-307X/42/8/080705},
year = {2025},
month = {aug},
publisher = {Chinese Physical Society and IOP Publishing Ltd},
volume = {42},
number = {8},
pages = {080705},
author = {Wang, Zhengxuan and Wu, Ruqian and Ma, Chunlan and Gong, Shijing and Zhao, Chuanxi and Zhang, Shuaikang and Wang, Guangtao and Wang, Tianxing and An, Yipeng},
title = {Symmetry-Constrained Anomalous Transport in the Altermagnetic Material CuX$_2$ (X = F, Cl)},
journal = {Chin. Phys. Lett.},
}

@Article{AM6,
AUTHOR = {Tamang, Rupam and Gurung, Shivraj and Rai, Dibya Prakash and Brahimi, Samy and Lounis, Samir},
TITLE = {Altermagnetism and Altermagnets: A Brief Review},
JOURNAL = {Magnetism},
VOLUME = {5},
YEAR = {2025},
NUMBER = {3},
pages = {17},
URL = {https://www.mdpi.com/2673-8724/5/3/17},
ISSN = {2673-8724},
DOI = {10.3390/magnetism5030017}
}

@article{zhou,
  title={Emergent multiferroic altermagnets and spin control via noncollinear molecular polarization},
  author={Zhu, Ziye and Liu, Yuntian and Duan, Xunkai and Zhang, Jiayong and Hao, Bowen and Wei, Su-Huai and {\v{Z}}uti{\'c}, Igor and Zhou, Tong},
  journal={Sci. China, Phys. Mech. Astron.},
  volume={68},
  number={12},
  pages={127562},
  year={2025},
  publisher={Springer}
}

@article{JACS,
author = {Ni, Xiaojuan and Ji, Huiwen and Liu, Feng and Br{\'e}das, Jean-Luc},
title = {Emergence of g-Wave Altermagnetism in Three-Dimensional Metal–Organic Frameworks},
journal = {J. Am. Chem. Soc.},
volume = {148},
number = {15},
pages = {15417-15425},
year = {2026},
doi = {10.1021/jacs.5c15427},
URL = { https://doi.org/10.1021/jacs.5c15427},
eprint = { https://doi.org/10.1021/jacs.5c15427}
}

@article{JACS2,
author = {Kashikar, Ravi and DeTellem, Derick and Ghosh, Partha Sarathi and Xu, Yixuan and Ma, Shengqian and Witanachchi, Sarath and Phan, Manh-Huong and Lisenkov, Sergey and Ponomareva, Inna},
title = {Coupling of the Structure and Magnetism to Spin Splitting in Hybrid Organic–Inorganic Perovskites},
journal = {J. Am. Chem. Soc.},
volume = {146},
number = {19},
pages = {13105-13112},
year = {2024},
doi = {10.1021/jacs.3c14744},
URL = { https://doi.org/10.1021/jacs.3c14744},
eprint = { https://doi.org/10.1021/jacs.3c14744}
}

@article{Nanote,
doi = {10.1088/1361-6528/ae6f22},
url = {https://doi.org/10.1088/1361-6528/ae6f22},
year = {2026},
month = {may},
publisher = {IOP Publishing},
volume = {37},
number = {22},
pages = {225702},
author = {Routh, Sayan and Patel, Shubham and Debnath, Tuhin and Mitra, Saikat and Nandy, Snehasish and Chowdhury, Avijit},
title = {Unravelling the origin of ferromagnetism and critical scaling behavior in orbital altermagnetic candidate CuBr$_2$ derived lead free Ruddlesden–Popper halide perovskite (PEA)$_2$CuBr$_4$},
journal = {Nanotechnology},
}

@article{HOIP1,
  title={Inversion of molecular chirality associated with ferroelectric switching in a high-temperature two-dimensional perovskite ferroelectric},
  author={Deng, Wen-Feng and Li, Yu-Xia and Zhao, Yan-Xin and Hu, Jie-Sheng and Yao, Zi-Shuo and Tao, Jun},
  journal={J. Am. Chem. Soc.},
  volume={145},
  number={9},
  pages={5545--5552},
  year={2023},
  publisher={ACS Publications}
}

@Article{HOIP2,
author ={Zang, Yipeng and Feng, Bolin and Gao, Xiaoqing},
title  ={Progress in chiral organic ferroelectrics},
journal  ={Chem. Commun.},
year  ={2026},
volume  ={62},
issue  ={1},
pages  ={31-44},
publisher  ={The Royal Society of Chemistry},
doi  ={10.1039/D5CC05018A},
url  ={http://dx.doi.org/10.1039/D5CC05018A},
}

@article{HOIP3,
  title={Chiral multiferroicity in two-dimensional hybrid organic-inorganic perovskites},
  author={Zheng, Haining and Ghosh, Arup and Swamynadhan, MJ and Zhang, Qihan and Wong, Walter PD and Wu, Zhenyue and Zhang, Rongrong and Chen, Jingsheng and Cimpoesu, Fanica and Ghosh, Saurabh},
  journal={Nat. Commun.},
  volume={15},
  number={1},
  pages={5556},
  year={2024},
  publisher={Nature Publishing Group UK London}
}

@article{CME1,
author = {Liu, Qi and Ren, Hui and Wei, Qi and Li, Mingjie},
title = {Multifunctional Chiral Halide Perovskites: Advancing Chiro-Optics, Chiro-Optoelectronics, and Spintronics},
journal = {Adv. Sci.},
volume = {12},
number = {34},
pages = {e09155},
doi = {https://doi.org/10.1002/advs.202509155},
url = {https://advanced.onlinelibrary.wiley.com/doi/abs/10.1002/advs.202509155},
eprint = {https://advanced.onlinelibrary.wiley.com/doi/pdf/10.1002/advs.202509155},
year = {2025}
}

@article{CME2,
   author = "Dong, Yifan and Hautzinger, Matthew P. and Haque, Md Azimul and Beard, Matthew C.",
   title = "Chirality-Induced Spin Selectivity in Hybrid Organic-Inorganic Perovskite Semiconductors", 
   journal= "Annu. Rev. Phys. Chem.",
   year = "2025",
   volume = "76",
   number = "Volume 76, 2025",
   pages = "519-537",
   doi = "https://doi.org/10.1146/annurev-physchem-082423-032933",
   url = "https://www.annualreviews.org/content/journals/10.1146/annurev-physchem-082423-032933",
   publisher = "Annual Reviews",
   issn = "1545-1593",
   type = "Journal Article",

  }

@article{CME3,
  title={Magneto-optical detection of photoinduced magnetism via chirality-induced spin selectivity in 2D chiral hybrid organic--inorganic perovskites},
  author={Huang, Zhengjie and Bloom, Brian P and Ni, Xiaojuan and Georgieva, Zheni N and Marciesky, Melissa and Vetter, Eric and Liu, Feng and Waldeck, David H and Sun, Dali},
  journal={ACS Nano},
  volume={14},
  number={8},
  pages={10370--10375},
  year={2020},
  publisher={ACS Publications}
}

@article{CME4,
author = {Evers, Ferdinand and Aharony, Amnon and Bar-Gill, Nir and Entin-Wohlman, Ora and Hedegård, Per and Hod, Oded and Jelinek, Pavel and Kamieniarz, Grzegorz and Lemeshko, Mikhail and Michaeli, Karen and Mujica, Vladimiro and Naaman, Ron and Paltiel, Yossi and Refaely-Abramson, Sivan and Tal, Oren and Thijssen, Jos and Thoss, Michael and van Ruitenbeek, Jan M. and Venkataraman, Latha and Waldeck, David H. and Yan, Binghai and Kronik, Leeor},
title = {Theory of Chirality Induced Spin Selectivity: Progress and Challenges},
journal = {Adv. Mater.},
volume = {34},
number = {13},
pages = {2106629},
doi = {https://doi.org/10.1002/adma.202106629},
url = {https://advanced.onlinelibrary.wiley.com/doi/abs/10.1002/adma.202106629},
year = {2022}
}

@article{MnCl4,
  title={Chiral weak ferromagnets formed in one-dimensional organic--inorganic hybrid manganese chloride hydrates},
  author={Taniguchi, Kouji and Huang, Po-Jung and Kimura, Shojiro and Miyasaka, Hitoshi},
  journal={Dalton Trans.},
  volume={51},
  number={44},
  pages={17030--17034},
  year={2022},
  publisher={Royal Society of Chemistry}
}

@article{2MnCl4,
author = {Asensio, Yaiza and Bahmani Jalali, Houman and Marras, Sergio and Gobbi, Marco and Casanova, Fèlix and Mateo-Alonso, Aurelio and Di Stasio, Francesco and Rivilla, Ivan and Hueso, Luis E. and Martín-García, Beatriz},
title = {Circularly Polarized Photoluminescence in Chiral Hybrid Organic–Inorganic Manganese Halide Perovskites: From Bulk Materials to Exfoliated Flakes},
journal = {Adv. Opt. Mater.},
volume = {12},
number = {21},
pages = {2400554},
doi = {https://doi.org/10.1002/adom.202400554},
url = {https://advanced.onlinelibrary.wiley.com/doi/abs/10.1002/adom.202400554},
eprint = {https://advanced.onlinelibrary.wiley.com/doi/pdf/10.1002/adom.202400554},
year = {2024}
}

@article{CASE1,
author = {Sun, Wei and Yang, Changhong and Wang, Wenxuan and Liu, Ying and Wang, Xiaotian and Huang, Shifeng and Cheng, Zhenxiang},
title = {Proposing Altermagnetic-Ferroelectric Type-III Multiferroics with Robust Magnetoelectric Coupling},
journal = {Adv. Mater.},
volume = {37},
number = {26},
pages = {2502575},
doi = {https://doi.org/10.1002/adma.202502575},
url = {https://advanced.onlinelibrary.wiley.com/doi/abs/10.1002/adma.202502575},
eprint = {https://advanced.onlinelibrary.wiley.com/doi/pdf/10.1002/adma.202502575},
year = {2025}
}

@article{CASE2,
  title = {Ferroelectric Switchable Altermagnetism},
  author = {Gu, Mingqiang and Liu, Yuntian and Zhu, Haiyuan and Yananose, Kunihiro and Chen, Xiaobing and Hu, Yongkang and Stroppa, Alessandro and Liu, Qihang},
  journal = {Phys. Rev. Lett.},
  volume = {134},
  issue = {10},
  pages = {106802},
  numpages = {7},
  year = {2025},
  month = {Mar},
  publisher = {American Physical Society},
  doi = {10.1103/PhysRevLett.134.106802},
  url = {https://link.aps.org/doi/10.1103/PhysRevLett.134.106802}
}

@article{CASE3,
author = {Bezzerga, Djamel and Khan, Imran and Hong, Jisang},
title = {High Performance Room Temperature Multiferroic Properties of w-MnSe Altermagnet},
journal = {Adv. Funct. Mater.},
volume = {35},
number = {43},
pages = {2505813},
doi = {https://doi.org/10.1002/adfm.202505813},
url = {https://advanced.onlinelibrary.wiley.com/doi/abs/10.1002/adfm.202505813},
eprint = {https://advanced.onlinelibrary.wiley.com/doi/pdf/10.1002/adfm.202505813},
year = {2025}
}

@article{CASE4,
  title = {Sliding Ferroelectric Control of Unconventional Magnetism in Stacked Bilayers},
  author = {Zhu, Yongqian and Gu, Mingqiang and Liu, Yuntian and Chen, Xiaobing and Li, Yuhui and Du, Shixuan and Liu, Qihang},
  journal = {Phys. Rev. Lett.},
  volume = {135},
  issue = {5},
  pages = {056801},
  numpages = {7},
  year = {2025},
  month = {Aug},
  publisher = {American Physical Society},
  doi = {10.1103/dmzg-ck2t},
  url = {https://link.aps.org/doi/10.1103/dmzg-ck2t}
}

@article{CASE5,
  title={Two-dimensional dual-switchable ferroelectric altermagnets: altering electrons and magnons},
  author={Wang, ShuaiYu and Wang, Wei-Wei and Fan, Jiaxuan and Zhou, Xiaodong and Li, Xiao-Ping and Wang, Lei},
  journal={Nano Lett.},
  volume={25},
  number={40},
  pages={14618--14624},
  year={2025},
  publisher={ACS Publications}
}

@article{MOKE,
author = "Gallego, Samuel V. and Etxebarria, Jesus and Elcoro, Luis and Tasci, Emre S. and Perez-Mato, J. Manuel",
title = "{Automatic calculation of symmetry-adapted tensors in magnetic and non-magnetic materials: a new tool of the Bilbao Crystallographic Server}",
journal = "Acta Crystallogr., Sect. A",
year = "2019",
volume = "75",
number = "3",
pages = "438--447",
month = "May",
doi = {10.1107/S2053273319001748},
url = {https://doi.org/10.1107/S2053273319001748},
}

@article{MOKE1,
  title = {Crystal chirality magneto-optical effects in collinear antiferromagnets},
  author = {Zhou, Xiaodong and Feng, Wanxiang and Yang, Xiuxian and Guo, Guang-Yu and Yao, Yugui},
  journal = {Phys. Rev. B},
  volume = {104},
  issue = {2},
  pages = {024401},
  numpages = {8},
  year = {2021},
  month = {Jul},
  publisher = {American Physical Society},
  doi = {10.1103/PhysRevB.104.024401},
  url = {https://link.aps.org/doi/10.1103/PhysRevB.104.024401}
}

@article{MOKE2,
  title = {Spin-order dependent anomalous Hall effect and magneto-optical effect in the noncollinear antiferromagnets ${\mathrm{Mn}}_{3}X\mathrm{N}$ with $X=\mathrm{Ga}$, Zn, Ag, or Ni},
  author = {Zhou, Xiaodong and Hanke, Jan-Philipp and Feng, Wanxiang and Li, Fei and Guo, Guang-Yu and Yao, Yugui and Bl\"ugel, Stefan and Mokrousov, Yuriy},
  journal = {Phys. Rev. B},
  volume = {99},
  issue = {10},
  pages = {104428},
  numpages = {13},
  year = {2019},
  month = {Mar},
  publisher = {American Physical Society},
  doi = {10.1103/PhysRevB.99.104428},
  url = {https://link.aps.org/doi/10.1103/PhysRevB.99.104428}
}

@article{MOKE3,
  title = {Pseudopotential-based first-principles approach to the magneto-optical Kerr effect: From metals to the inclusion of local fields and excitonic effects},
  author = {Sangalli, Davide and Marini, Andrea and Debernardi, Alberto},
  journal = {Phys. Rev. B},
  volume = {86},
  issue = {12},
  pages = {125139},
  numpages = {8},
  year = {2012},
  month = {Sep},
  publisher = {American Physical Society},
  doi = {10.1103/PhysRevB.86.125139},
  url = {https://link.aps.org/doi/10.1103/PhysRevB.86.125139}
}

@article{MOKE4,
  title = {Polar magneto-optical Kerr effect for low-symmetric ferromagnets},
  author = {Rathgen, Helmut and Katsnelson, Mikhail I. and Eriksson, Olle and Zwicknagl, Gertrud},
  journal = {Phys. Rev. B},
  volume = {72},
  issue = {1},
  pages = {014451},
  numpages = {13},
  year = {2005},
  month = {Jul},
  publisher = {American Physical Society},
  doi = {10.1103/PhysRevB.72.014451},
  url = {https://link.aps.org/doi/10.1103/PhysRevB.72.014451}
}

@article{SSG1,
  title = {Spin-Group Symmetry in Magnetic Materials with Negligible Spin-Orbit Coupling},
  author = {Liu, Pengfei and Li, Jiayu and Han, Jingzhi and Wan, Xiangang and Liu, Qihang},
  journal = {Phys. Rev. X},
  volume = {12},
  issue = {2},
  pages = {021016},
  numpages = {19},
  year = {2022},
  month = {Apr},
  publisher = {American Physical Society},
  doi = {10.1103/PhysRevX.12.021016},
  url = {https://link.aps.org/doi/10.1103/PhysRevX.12.021016}
}

@article{SSG2,
  title = {Spin Space Groups: Full Classification and Applications},
  author = {Xiao, Zhenyu and Zhao, Jianzhou and Li, Yanqi and Shindou, Ryuichi and Song, Zhi-Da},
  journal = {Phys. Rev. X},
  volume = {14},
  issue = {3},
  pages = {031037},
  numpages = {33},
  year = {2024},
  month = {Aug},
  publisher = {American Physical Society},
  doi = {10.1103/PhysRevX.14.031037},
  url = {https://link.aps.org/doi/10.1103/PhysRevX.14.031037}
}

@article{SSG3,
  title = {Enumeration and Representation Theory of Spin Space Groups},
  author = {Chen, Xiaobing and Ren, Jun and Zhu, Yanzhou and Yu, Yutong and Zhang, Ao and Liu, Pengfei and Li, Jiayu and Liu, Yuntian and Li, Caiheng and Liu, Qihang},
  journal = {Phys. Rev. X},
  volume = {14},
  issue = {3},
  pages = {031038},
  numpages = {33},
  year = {2024},
  month = {Aug},
  publisher = {American Physical Society},
  doi = {10.1103/PhysRevX.14.031038},
  url = {https://link.aps.org/doi/10.1103/PhysRevX.14.031038}
}

@article{x-wave,
doi = {10.35848/1882-0786/ae4311},
url = {https://doi.org/10.35848/1882-0786/ae4311},
year = {2026},
month = {mar},
publisher = {IOP Publishing},
volume = {19},
number = {3},
pages = {030101},
author = {Ezawa, Motohiko},
title = {Quantum geometry and X-wave magnets with X = p, d, f, g, i},
journal = {Appl. Phys. Express},
}

@Article{D-case1,
AUTHOR = {Li, Renfu and Jiang, Lulu and Zou, Qinghua and Bai, Jianlong and Wu, Lingkun and Li, Jianrong and Liao, Jinsheng},
TITLE = {Highly Luminescent and Scintillating Hybrid Halide of (C$_{13}$H$_{25}$N)$_2$[MnBr$_4$] Enabled by Rigid Cation},
JOURNAL = {Molecules},
VOLUME = {30},
YEAR = {2025},
NUMBER = {10},
pages = {2157},
URL = {https://www.mdpi.com/1420-3049/30/10/2157},
PubMedID = {40430329},
ISSN = {1420-3049},
DOI = {10.3390/molecules30102157}
}

@article{D-case2,
author = {Yu, Fang and Li, Shu-Yao and Yang, Hai-Rong and Shen, Jie and Yin, Ming-Xia and Tian, Yan-Rui and Zhang, Ya-Tong and Kong, Xiang-Wen and Lei, Xiao-Wu},
title = {Crystal-Rigidifying Strategy in Hybrid Manganese Halide to Achieve Narrow Green Emission and High Structural Stability},
journal = {Inorg. Chem.},
volume = {63},
number = {30},
pages = {14116-14125},
year = {2024},
doi = {10.1021/acs.inorgchem.4c01953},
URL = { https://doi.org/10.1021/acs.inorgchem.4c01953
},
eprint = {https://doi.org/10.1021/acs.inorgchem.4c01953}
}

@Article{D-case3,
author ={Wang, Mengzhu and Wang, Xiaoming and Zhang, Bintao and Li, Feiyang and Meng, Haixing and Liu, Shujuan and Zhao, Qiang},
title  ={Chiral hybrid manganese(ii) halide clusters with circularly polarized luminescence for X-ray imaging},
journal  ={J. Mater. Chem. C},
year  ={2023},
volume  ={11},
issue  ={9},
pages  ={3206-3212},
publisher  ={The Royal Society of Chemistry},
doi  ={10.1039/D2TC05379A},
url  ={http://dx.doi.org/10.1039/D2TC05379A},
}

@article{D-case4,
author = {Zhang, Wei and Zheng, Wei and Li, Lingyun and Huang, Ping and Xu, Jin and Zhang, Wen and Shao, Zhiqing and Chen, Xueyuan},
title = {Unlocking the Potential of Organic-Inorganic Hybrid Manganese Halides for Advanced Optoelectronic Applications},
journal = {Adv. Mater.},
volume = {36},
number = {39},
pages = {2408777},
doi = {https://doi.org/10.1002/adma.202408777},
url = {https://advanced.onlinelibrary.wiley.com/doi/abs/10.1002/adma.202408777},
eprint = {https://advanced.onlinelibrary.wiley.com/doi/pdf/10.1002/adma.202408777},
year = {2024}
}

@article{D-case5,
title = {Environmental-friendly lead-free chiral Mn-based metal halides with efficient circularly polarized photoluminescence at room temperature},
journal = {J. Alloys Compd.},
volume = {910},
pages = {164892},
year = {2022},
issn = {0925-8388},
doi = {https://doi.org/10.1016/j.jallcom.2022.164892},
url = {https://www.sciencedirect.com/science/article/pii/S092583882201283X},
author = {Beibei Wang and Chao Wang and Ya Chu and Haoyue Zhang and Mengjiao Sun and Hui Wang and Shiping Wang and Guangjiu Zhao},
}

@article{D-case6,
author = {Li, Jing and Luo, Qiulian and Wei, Jianwu and Zhou, Liya and Chen, Peican and Luo, Binbin and Chen, Yibo and Pang, Qi and Zhang, Jin Zhong},
title = {Circularly Polarized Luminescence Induced by Hydrogen-Bonding Networks in a One-Dimensional Hybrid Manganese(II) Chloride},
journal = {Angew. Chem. Int. Ed.},
volume = {63},
number = {24},
pages = {e202405310},
doi = {https://doi.org/10.1002/anie.202405310},
url = {https://onlinelibrary.wiley.com/doi/abs/10.1002/anie.202405310},
eprint = {https://onlinelibrary.wiley.com/doi/pdf/10.1002/anie.202405310},
year = {2024}
}

@article{D-case7,
author = {Ding, Zijin and Chen, Quanlin and Jiang, Yuanzhi and Yuan, Mingjian},
title = {Structure-Guided Approaches for Enhanced Spin-Splitting in Chiral Perovskite},
journal = {JACS Au},
volume = {4},
number = {4},
pages = {1263-1277},
year = {2024},
doi = {10.1021/jacsau.3c00835},
URL = { https://doi.org/10.1021/jacsau.3c00835},
eprint = {  https://doi.org/10.1021/jacsau.3c00835}
}

\end{document}